УДК378.(4:6):377.8]+372.851]:004

**MaiiaV. Popel**
PhD of Pedagogical Sciences, Senior Researcher
Institute of Information Technologies and Learning Tools of NAES of Ukraine, Kyiv, Ukraine
ORCID ID 0000-0002-8087-962X
*popelmaya@gmail.com*

# USING COCALC AS A TRAINING TOOL FOR MATHEMATICS TEACHERS' PRE-SERVICE TRAINING

**Abstract.** This paper deals with the problem of theoretical justification and development of scientific and methodological support for using the cloud service CoCalc as a tool for the formation of mathematics teachers' professional competencies. The following contradictions exist concerning the processes of forming these competencies: between the level of abstraction of mathematical objects and the possibilities of providing their visualization by computer interpretation; between the expediency of widespread use of ICT services for mathematical purposes in the training of pre-service teachers of mathematics and the lack of opportunities for their provision to ICT subdivisions and pedagogical universities; between the feasibility of using ICT outsourcing of cloud infrastructure for the pre-service training of mathematics teachers in Ukraine and the non-adaptability of foreign cloud-based mathematical services to the requirements of national educational standards; between the possibilities of application of cloud mathematical services in the process of formation of professional competencies of mathematics teachers and the absence of an adequate method for their implementation. The paper describes the professional training of mathematics teachers in universities of Ukraine, and considers the national and foreign experience of using the cloud-based services in mathematics teachers' pre-service training and also the tendencies and prospects of using CoCalc in teaching mathematical disciplines. The process of system design of mathematics teachers' professional competencies is characterized, and the model of using the cloud service CoCalc as a tool for forming mathematics teachers' professional competencies is developed. The indicators and levels (high, sufficient, medium, low) were identified for each component of the pre-service mathematics teachers' professional competence system within the proposed model. The method of using CoCalc as a tool for forming professional competencies of mathematics teachers is developed and its basic components such as purpose, content, tools, methods and results are elaborated. Information regarding the stages of research and also the experimental work objectives and content are presented; the quantitative and qualitative analysis of the main stages (ascertaining, formative) of the pedagogical experiment is performed, confirming the hypothesis of the study.

## 1. INTRODUCTION

**The problem statement.** Improvement of content and components of courses of mathematical disciplines and methods of their teaching are among the key issues for improving the quality of professional training, especially at the pedagogical university in the study of many disciplines (such as differential geometry and topology, differential equations, probability theory and others). The students have considerable difficulties mastering abstract mathematical concepts. One possible way to handle them is the use of visual interpretations of mathematical concepts and assertions. Significant didactic opportunities for implementing the principle of visibility arise with the use of ICT in learning.

In this regard, revealing the prospects of using cloud services for teaching mathematical disciplines, their role and place within the organization of the educational process, and methodological principles of their application are among the topical problems of the theory and methodology of ICT in education. Essential to their solution is the scientific and methodological justification of using leading cloud services, particularly CoCalc, for mathematical purposes.

The research problem is the theoretical substantiation and development of scientific and methodological support for using the cloud service CoCalc as a tool for forming mathematics teachers' professional competencies.

**Analysis of recent research and publications.** The problems of mathematics teachers' training in domestic higher educational institutions are considered in the work of leading researchers: I. A. Akulenko[1], V. H. Bevz, M. I. Zhaldak [2], I. V. Lovyanova [3], H. O. Mykhalin [4], N. V. Morze [5], T. O. Oliynyk, M. V. Prats′ovytyy [6], S. A. Rakov, Y. S. Rams′kyy [7], O. I. Skafa, Z. I. Slyepkan′, O. V. Spivakovs′kyy [8], Y. V. Tryus [9], V. O. Shvets′ and others.

A separate group is the research of T. L. Arkhipova [10], N. V. Bakhmat, V. Y. Bykov, D. Blank [11], T. V. Zaytseva [10], U. P. Kohut, Y. H. Lotyuk [12], J. Marshall [11], N. V. Morze [13], V. P. Oleksyuk [14], K. J. O'Hara [11], K. I. Slovak, S. V. Shokalyuk and others, devoted to the application of cloud services in the process of training mathematics teachers. Significant contribution to the analysis of educational and research opportunities of the use of the cloud-based ICT was made by such scientists as: M. Armbrust [15], R. Griffith [15], M. Miller [16], K. Subramanian [17], N. Sultan [18], P. Thomas [19], A. Fox [15], Y. Khmelevsky[20], W. Chang[21].

Still, the problems of scientific and methodological support of the process of using different types of cloud services require further research following the current trends of the mathematical software development and application in the learning process.

The following contradictions exist concerning the processes of forming these competencies:
− between the level of abstraction of mathematical objects and the possibilities of providing their visualization by computer interpretation;
− between the expediency of widespread use of ICT services for mathematical purposes in the training of pre-service teachers of mathematics and the lack of opportunities for their provision by ICT subdivisions of pedagogical universities;
− between the feasibility of using ICT outsourcing of cloud infrastructure for the pre-service training of mathematics teachers in Ukraine and the non-adaptability of foreign cloud-based mathematical services to the requirements of national educational standards;
− between the possibilities of application of cloud mathematical services in the process of formation of professional competencies of mathematics teachers and the absence of an adequate method for their implementation.

**The purpose of the article.** The purpose of the research is to substantiate and develop the method of using the cloud service CoCalc as a tool for the formation of professional competencies of a teacher of mathematics.

In accordance with the purpose of the study, the following tasks are set:
− to examine the state of the problem of using cloud services in mathematics teachers' training;
− to identify the professional competencies that can be gained using CoCalc, and to determine the indicators and levels of their formation;
− to develop the model for using the CoCalc cloud service as a tool for the formation of mathematics teachers' professional competencies;
− to develop the methodology of using the CoCalc cloud service as a tool for the formation of mathematics teachers' professional competencies and to verify experimentally its effectiveness.

To solve these tasks, the following methods and sources have been used, in particular: analysis, generalization, systematization of scientific and methodological sources on the research problem, analysis of modern cloud services to determine the theoretical foundations, Internet resources, software in order to substantiate the components of the model of using the cloud service CoCalc as a tool for forming the professional competencies of a mathematics teacher; diagnostic techniques (purposeful pedagogical observations, interviews with teachers and students, questionnaires, testing, analysis of the experience of teachers in the main provisions of the study) to determine the state of the research problem; pedagogical experiments to verify the method of using the cloud service CoCalc as a tool for forming professional competence of a mathematics teacher; and statistics, that is, implementation of calculations for the quantitative and qualitative analysis of learning outcomes based on the developed methodology.

## 2. RESEARCH RESULTS

### 2.1. Theoretical foundations for using cloud services in the training of mathematics teachers

The research substantiates the key concept during the study: professional competence of a mathematics teacher – an individual's ability to exercise professional activities in teaching mathematics to students based on knowledge, abilities, skills and personal attitudes, and to achieve certain results.

The paper considers the competence of mathematics trainee teachers: general professional and special professional. According to the results of the study, the special professional competencies that are appropriate to the formation with the use of cloud services were identified as follows:
– the ability to use professionally profiled knowledge in the field of mathematics for the statistical processing of experimental data and mathematical modelling of natural phenomena and processes;
– the ability to use mathematical devices to simulate a variety of processes;
– the ability to work with a computer at the level of the user and the specialist in the field of ICT.

In the process of researching domestic and foreign experience, the following benefits of using cloud services for mathematical purposes have been identified:
– saving resources (reducing the load on the environment, costs for the acquisition and modernization of computer equipment, personnel payment);
– mobility of access (classes with the assimilation of material at a convenient time and in a convenient place);
– elasticity (providing additional computational resources at the request of the user).

The application of cloud services leads to the emergence and development of forms of training organization focused on joint educational activities on the Internet. It is shown that cloud services in the training of pre-service mathematics teachers are expediently used as tools for:
– communication (synchronous – chats, voice and video; asynchronous – mail, forums),
– collaboration (data access, sharing and collaboration with other users),
– storage and processing of data.

The trends for using CoCalc in the training of mathematics teachers-to-be are as follows:
– organization of educational communication;

- support of individual and group forms of organization of educational activities (classroom and extracurricular);
- support of training management;
- providing visibility by constructing different interpretations of mathematical models, visualizing mathematical abstractions, etc.;
- providing accessibility and knowledge using the shared interface for access to environmental objects and reliable open source software;
- increase in time and spatial mobility; the formation of a single learning environment, the content of which develops in the learning process.

### 2.2. Modelling the process of using CoCalc cloud service as a tool for forming mathematics teachers' professional competencies

The main ideas of the study are reflected in the hypothesis: methodologically grounded use of the cloud service CoCalc in teaching mathematical disciplines will increase the level of formation of mathematics teachers' professional competencies.

Summarizing the system of professional competencies of a teacher of mathematics (constructed on the basis of the system proposed by M. I. Zhaldak, Y. S. Ramsky, M. V. Rafalskaya, as well as that of O. M. Spirin) and the results of the study, the following components of the system of professional competencies of mathematics trainee teachers were revealed:

1. general professional competence;
2. special professional competence:
- subjective competencies (scientific, subject-pedagogical),
- technological competencies (informational-methodological, informational-technological),
- professional-practical competence (mathematical, methodological).

Each component of the proposed system of professional competencies is characterized by four levels of their formation: high, sufficient, average, and low.

On the basis of the designed system of professional competencies of mathematics teachers [22], the model for using the CoCalc cloud service as a tool for forming the professional competencies of mathematics teachers was developed (Fig. 1).

When creating the model for forming the professional competencies of mathematics teachers, the following principles were considered [23]:
- the principle of science orientation;
- professional orientation;
- self-realization of interdisciplinary integration;
- variability.

The model consists of the following components:
- target;
- stimulating and motivational;
- content;
- operational-activity;
- valuation-regulatory.

The model covers three stages of the formation of professional competencies: propaedeutic, forming and developmental.

The basis of the model is the goal, which is determined by the need of the society in the training of a competent teacher of mathematics, within the competence approach to learning and the trends of using ICT outsourcing in the learning and development of ICT-based tools.

The development of ICT contributes to the emergence of the cloud-based learning services, the study of which, in turn, is a part of the teaching of mathematical disciplines. Their content is compiled by semester and covers four courses for pre-service teachers of mathematics. The content contains a list of basic mathematical disciplines and domain-oriented learning of how to use the cloud service CoCalc. Each of the stages of implementation of the model is closely linked to the basic mathematical disciplines studied by undergraduate students.

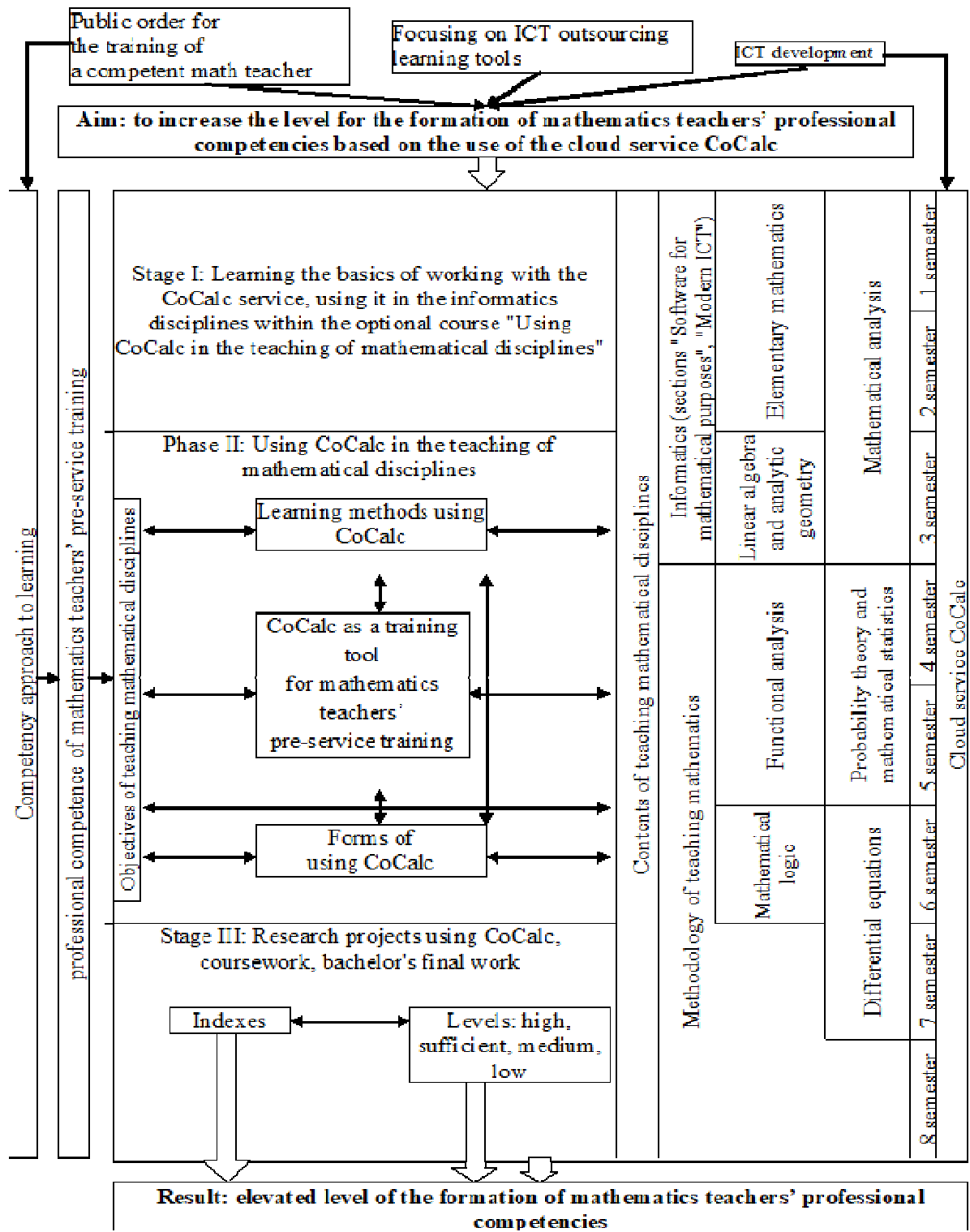

*Fig. 1. The use of the CoCalc cloud service as a tool of the mathematics teacher's professional competencies formation*

### 2.3. The method of using CoCalc as a training tool for mathematics teachers' pre-service training

The proposed method involves two stages of the introduction of CoCalc into the learning process [23]:

1. Teaching bythe optional program "Using CoCalc in the process of studying mathematical disciplines" as an element of the content of training, retraining, professional development of scientific and scientific-pedagogicalpersonnel.

2. Introduction of the system of trainings, seminars, webinars, individual consultations which can be carried out during the pilot experimental study (project) on the deployment of a cloud-based information and educational environment at an educational institution.

At the first stage of the formation of professional competencies according to the proposed method it is expedient to include in the variation part of the educational and professional training program of Bachelor of Mathematics an optional course "Use of CoCalc in the process of studying mathematical disciplines". The learning process is aimed at considering interdisciplinary relations (mathematical and informatics disciplines).

To work with CoCalc the students must have the following skills that are acquired at the propaedeutic stage:
- to register and authorize in the system;
- to create educational resources; work with sagews sheets (including the most common modes, know the basics of languages: LaTeX, Python, HTML);
- to embed video, audio, questionnaires, graphic files into sagews resource;
- to chat in learning resources chats and in the sage-chat type resource;
- to work with an educational resource like tex;
- to download new resources from electronic media.

At the second stage the formation of professional competencies takes place within the framework of the study of normative mathematical educational disciplines.

Phase III covers the implementation of research and development projects using CoCalc, coursework, graduation thesisfor theBachelor's degree. The main objective of the method of using cloud service CoCalc as a tool for mathematics teachers' professional competencies formation is to increase the level of their professional competencies.

### 2.4. Organization and results of the experimental research

The control and experimental groups were formed as follows: the control groups (CG) included the students trained according to the traditional method of mathematics teachers' professional competencies formation; the experimental groups (EG) included students trained according to the author's technique for using CoCalc as a training tool for mathematics teachers' pre-service training.

Summarizing the obtained results of the confirmatory stage of the pedagogical experiment, it can be argued that:
- the vast majority of students and teachers have the opportunity to work with the cloud-based CoCalc service both at universities and at home;
- teachers in most cases do not use cloud services in the learning process, except for their use as a cloud storage;
- teachers are interested in implementing the CoCalc cloud service in the learning process, but students are not ready for this;
- students at the beginning of the experiment showed a low level of information and technological and subject-pedagogical competencies formation, sufficient mathematical competencies;

- students and teachers use only free software tools (mostly local computer mathematics systems).

*Table 1.*
**Comparison of the distribution of the experimental and the control groups of students after the formation stage of the experiment**

| Scale levels | Components | | | | | |
|---|---|---|---|---|---|---|
| | Informational and technological | | Mathematical | | Subject-pedagogical | |
| | KG | EG | KG | EG | KG | EG |
| At the initial stage of the experiment | | | | | | |
| High | 2 | 2 | 15 | 15 | 12 | 12 |
| Sufficient | 12 | 14 | 22 | 23 | 6 | 10 |
| Average | 25 | 25 | 11 | 12 | 7 | 12 |
| Low | 22 | 20 | 13 | 11 | 36 | 27 |
| Total | 61 | 61 | 61 | 61 | 61 | 61 |
| The empirical value of Fisher's criterion | 0,49 | | 0,19 | | 0,77 | |
| The critical value of Fisher's criterion | 1,64 | | 1,64 | | 1,64 | |
| After forming stage of the experiment | | | | | | |
| High | 2 | 5 | 8 | 15 | 11 | 16 |
| Sufficient | 12 | 28 | 18 | 25 | 6 | 13 |
| Average | 28 | 21 | 20 | 16 | 9 | 14 |
| Low | 19 | 7 | 15 | 5 | 35 | 18 |
| Total | 61 | 61 | 61 | 61 | 61 | 61 |
| The empirical value of Fisher's criterion | 3,61 | | 2,57 | | 2,26 | |
| The critical value of Fisher's criterion | 1,64 | | 1,64 | | 1,64 | |

The following components of the subject, technological and professional-practical competencies were examined: subject-pedagogical, informational-technological and mathematical competencies. Each component was considered separately, and the values were calculated according to the levels: high, sufficient, average and low. For data analysis, matches (at the initial stage of the experiment) and differences (after the forming stage of the experiment) of the experimental and control group characteristics (Table 1) were determined according to Fisher's criterion. For this purpose, statistical hypotheses were formulated: the absence of differences between the levels of formation of the individual components of the system of professional competencies and the significance of differences between the levels of formation of selected components.

Analyzing the obtained results at the summarising stage of the experiment, it can be concluded that the levels of formation of professional competencesof mathematics trainee teachers in control and experimental groups coincide with the level of significance $\alpha=0,05$.

Comparing the levels of the formation of professional competencies in the control and experimental groups at the beginning of the formative stage and at the end of the experiment, one can observe an increase in the proportion of students with high and average levels of professional competence.

The analysis of the results of the forming stage of the pedagogical experiment showed that the distribution of the levels of the formation of professional competencies in the experimental and control groups of mathematics trainee teachers has statistically significant differences due to the implementation of the developed method of using the cloud service CoCalc, which confirms the hypothesis of the study.

## 3. CONCLUSIONS AND PROSPECTS FOR FURTHER RESEARCH

The analysis of the national and foreign experience of the state-of-the-art cloud services introduction into the educational process showed that there are the cloud versions of various systems of computer mathematics that give rise to trends in the development of software for mathematical purposes. There are the tendencies of more active use of the cloud-based platforms for the software delivery, wider use of services virtualization, as well as their delivery as a service.

On the basis of the proposed research the model of using CoCalc cloud service as a tool of forming professional competencies of the mathematics teachers was approved, taking into account the links between the components of professional competencies and all cycles of the disciplines of the mathematics teachers' training programs for pedagogical universities. The use of the cloud service CoCalc proved to be effective at three stages of the development of professional competencies. It has been found out that the use of this cloud service in the process of pre-service training of mathematics teachers affects first of all the formation of special professional competencies.

As a part of the methodology of using the cloud service CoCalc as a tool for forming the professional competences of the teachers of mathematics, the interrelated goal, content, forms of organization, methods and tools of learning were revealed.

Among the forms of learning organization using cloud service CoCalc there are such as: dialog forms, individual and group consultations, independent work, practical work, individual work, in-pair work, frontal-collective work, collective and individual projects. The implementation of the methodology is effective if it is performed in three stages: stage I – propedeutic, stage II – formative, stage III – developmental. It has been experimentally confirmed that the level of formation of professional competences of pre-service mathematics teachers will be higher if in the process of teaching the developed method of using the cloud service CoCalc is pedagogically grounded and introduced.

The results of the pedagogical experiment, checked using the Fisher and Wilcoxon-Mann-Whitney criteria, gave reason to believe that the hypothesis of the study was confirmed.

The study does not exhaust all aspects of the problem. The continuation of research on this issue is advisable in the following areas: the development of theoretical and methodological principles for designing a cloud-based environment for the teaching of mathematical disciplines to mathematics teachers-to-be at a pedagogical university; the development of the method of using cloud service CoCalc in the process of improving the skills of teachers of mathematics.

# ХМАРНИЙ СЕРВІС COCALC ЯК ЗАСІБ ФОРМУВАННЯ ПРОФЕСІЙНИХ КОМПЕТЕНТНОСТЕЙ ВЧИТЕЛЯ МАТЕМАТИКИ

**Попель Майя Володимирівна**
кандидат педагогічних наук, старший науковий співробітник
Інститут інформаційних технологій і засобів навчання НАПН України, м.Київ, Україна
ORCID ID 0000-0002-8087-962X
*popelmaya@gmail.com*

**Анотація.** У роботі досліджується проблема теоретичного обґрунтування та розроблення науково-методичного супроводу процесу використання хмарного сервісу CoCalc як засобу формування професійних компетентностей учителя математики.У підході до вивчення цієї проблеми спостерігаються такі суперечності: між рівнем абстракції математичних об'єктів і можливостями забезпечення їх візуалізації шляхом комп'ютерної інтерпретації; між доцільністю широкого використання ІКТ-сервісів математичного призначення у підготовці майбутніх учителів математики та недостатніми можливостями їх забезпечення ІКТ-підрозділами педагогічних ЗВО; між доцільністю використання ІКТ-аутсорсингу хмарної інфраструктури навчання майбутніх учителів математики в Україні та неадаптованістю зарубіжних хмарних математичних сервісів до вимог вітчизняних освітніх стандартів; між можливостями застосування хмарних математичних сервісів у процесі формування професійних компетентностей учителя математики та відсутністю розробленої відповідної методики їх упровадження. Вивчено стан професійної підготовки вчителів математики у ЗВО України, проаналізовано вітчизняний і зарубіжний досвід використання хмарних сервісів у навчанні майбутніх учителів математики, виявлено тенденції та напрями використання CoCalcу навчанні математичних дисциплін.Охарактеризовано процес проектування системи професійних компетентностей учителя математики, розроблено модель використання хмарного сервісу CoCalc як засобу формування професійних компетентностей учителя математики. Для кожного складника системи професійних компетентностей майбутнього вчителя математики були визначені показники сформованості професійних компетентностей та відповідні їм рівні (високий, достатній, середній, низький).Обґрунтовано методику використання CoCalc як засобу формування професійних компетентностей учителя математики та розроблено її основні компоненти: мету, зміст, засоби, методи і форми використання цього хмарного сервісу, результат.Наведено відомості щодо етапів дослідження, завдань та змісту експериментальної роботи, виконано кількісне та якісне опрацювання результатів констатувального, формувального та підсумкового етапів педагогічного експерименту, що підтвердили гіпотезу дослідження.

**Ключові слова:** вчителі математики; математичні дисципліни; професійні компетентності; хмарні технології; хмарні сервіси;CoCalc.

# ОБЛАЧНЫЙ СЕРВИС COCALC КАК СРЕДСТВО ФОРМИРОВАНИЯ ПРОФЕССИОНАЛЬНЫХ КОМПЕТЕНТНОСТЕЙ УЧИТЕЛЯ МАТЕМАТИКИ

**Попель Майя Владимировна**
кандидат педагогических наук, старший научный сотрудник
Институт информационных технологий и средств обучения НАПН Украины, г. Киев, Украина
ORCID ID0000-0002-8087-962X
*popelmaya@gmail.com*

**Аннотация.** В работе исследуется проблема теоретического обоснования и разработки научно-методического сопровождения процесса использования облачного сервиса CoCalc как средства формирования профессиональных компетентностей учителя математики. В подходе к изучению этой проблемы наблюдаются следующие противоречия: между уровнем абстракции математических объектов и возможностями обеспечения их визуализации путем компьютерной интерпретации; между целесообразностью широкого использования ИКТ-сервисов математического назначения в подготовке будущих учителей математики и недостаточными возможностями их обеспечения ИКТ-подразделениями педагогических вузов; между целесообразностью использования ИКТ-аутсорсинга облачной инфраструктуры обучения будущих учителей математики в Украине и неадаптированностью зарубежных облачных математических сервисов к требованиям отечественных образовательных стандартов; между возможностями применения облачных математических сервисов в процессе формирования профессиональных компетентностей учителя математики и отсутствием разработок соответствующей методики их применения. Изучено состояние профессиональной подготовки учителей математики в вузах Украины, проанализированы отечественный и зарубежный опыт использования облачных сервисов в обучении будущих учителей математики, выявлены тенденции и направления использования CoCalc в обучении математическим дисциплинам. Охарактеризован процесс проектирования системы профессиональных компетентностей учителя математики, разработана модель использования облачного сервиса CoCalc как средства формирования профессиональных компетентностей учителя математики. Для каждого компонента системы профессиональных компетентностей будущего учителя математики были определены показатели сформированности профессиональных компетентностей и соответствующие им уровни (высокий, достаточный, средний, низкий). Обоснована методика использования CoCalc как средства формирования профессиональных компетентностей учителя математики и разработаны ее основные компоненты: цель, содержание, средства, методы и формы использования этого облачного сервиса, результат. Приведены сведения по этапам исследования, задачи и содержание экспериментальной работы, выполнена количественная и качественная обработка результатов констатирующего, формирующего и заключительного этапов педагогического эксперимента, подтверждена гипотеза исследования.

**Ключевые слова:** учителя математики; математические дисциплины; профессиональные компетентности; облачные технологии; облачные сервисы; CoCalc.